\documentclass{aa}  
\defcitealias{sanroman}{SR15}
\usepackage{xspace}

\newcommand{\gc}{GCs\xspace}
\newcommand{\km}{km s$^{-1}$\xspace}

\newcommand{\teff}{T$_{\rm eff}$\xspace}
\newcommand{\logg}{$\log g$\xspace}
\newcommand{\vt}{v$_{\rm t}$\xspace}
\newcommand{\feh}{[Fe/H]\xspace}

\usepackage{threeparttable}
\usepackage{graphicx}
\usepackage{stackengine}	
\usepackage{amsmath}	
\usepackage{amssymb}	
\usepackage{hyperref}
\hypersetup{
    colorlinks=true,
    citecolor=blue,
    linkcolor=blue,
    filecolor=blue,      
    urlcolor=blue,
    }
\usepackage{txfonts}

\begin{document}

   \title{Hint of bimodal Mg-Al anticorrelation in the metal-poor Globular Cluster NGC 4372}


   \author{Camila Beltrand\inst{1,2}, Macarena Uribe\inst{3}, Sandro Villanova\inst{3} \& C. Mu\~noz\inst{1}}

   \institute{Departamento de Astronom\'ia, Universidad de La Serena, Av. Ra\'ul Bitr\'an 1305, La Serena, Chile\\
    \email{camila.beltrand@userena.cl}
    \and Facultad de Ingeniería y Arquitectura, Universidad Central de Chile, Avenida Francisco de Aguirre 0405, La Serena, Chile
   \and Universidad Andres Bello, Facultad de Ciencias Exactas, Departamento de F{\'i}sica y Astronom{\'i}a - Instituto de Astrof{\'i}sica, Autopista Concepci\'on-Talcahuano 7100, Talcahuano, Chile
  }

   \date{Received XXXX; accepted XXXX}

 \titlerunning{Hint of bimodal  Mg-Al in the Globular Cluster NGC 4372}
\authorrunning{Camila Beltrand et al.}
  \abstract
{We present a detailed chemical abundance analysis of seven red giant branch stars in the very metal-poor globular cluster NGC~4372, based on high-resolution UVES spectra. Our study aims to characterize the chemical evolution of NGC~4372 and investigate the presence of multiple populations. We derived abundances for 17 elements, including light, iron-peak, $\alpha$, and heavy elements, using both equivalent width and spectral synthesis methods. We find a mean cluster metallicity of [Fe/H]=$-$2.36 $\pm$ 0.03 dex, with no evidence of an intrinsic iron spread. The cluster exhibits a typical $\alpha$-enhancement of [$\alpha$/Fe]= 0.33 $\pm$ 0.05 dex and a well defined Na-O anticorrelation, confirming that NGC 4372 hosts at least two distinct stellar populations. Furthermore, the measured ratio of [Ba/Eu] =$-0.47 \pm 0.04$ indicates that the synthesis of heavy elements was dominated by $r$-process, consistent with other very metal-poor environments. We also identify a potential bimodal Mg-Al anticorrelation, which aligns with the patterns observed in GCs of similar mass and metallicity. These results reinforce the status of NGC 4372 as a cornerstone for understanding the early chemical enrichment of the Galactic Halo.}


   \maketitle
%
\section{Introduction}

Globular Clusters (GCs) are ideal witnesses of past star formation and chemical evolution of the Galaxy. They are among the oldest known objects in the universe and serve as natural laboratories for studying a wide variety of fundamental problems in galactic astrophysics.

Today, the presence of multiple populations (MPs) within GCs is extensively associated with star-to-star spread of light elements \citep[from C to Si;][]{rev2004,bastianlardo18} and with the detection of multiple main sequences and giant branches in GCs \citep{piotto2012,piotto2015}.
This spread in light-elements is most likely due to early self-pollution within the cluster after its formation, allowing for the birth of two or more generations of stars \citep[see, e.g.,][and references therein]{rev2004,rev2012, bastianlardo18} separated by a few tens to hundreds of Myr. In this scenario, subsequent generations of stars are born from material enriched by the older, first generation (FG) stars. FG stars exhibit a primordial component similar to that of Milky Way field stars with the same metallicity. 

The most common signature of these abundance variations is the Na$-$O anticorrelation \citep{2009a}, traditionally considered the principal chemical signature to define a Galactic GC \citep{carretta2010}. However, this paradigm has been challenged by an increasing number of exceptions. Clusters such as Ruprecht~106 \citep{rup106,dotter18}, Palomar 12 \citep{cohen04}, and the recently characterized Patchick 126 \citep{garro26} show no evidence of the Na$-$O anticorrelation. Furthermore, 'Type II' clusters like NGC 1261 represent a more complex sub-population of GCs characterized by an enrichment in metallicity and $s-$process elements, often appearing as distinct sequences in multi-filter photometry \citep{milone17}. These features further complicate the standard classification of these systems \citep{munoz21}.

Several kinds of polluters for the light-elements have been proposed: intermediate-mass asymptotic giant branch (AGB) stars \citep{ventura2001,dercole2008}, fast-rotating massive stars \citep{decressin2007,krause2013}, very massive (VMSs; $\geq$ 5000 M$\odot$) stars \citep{Denissenkov_2014}, and massive interacting binaries \citep{demink2009}.

On the other hand, some massive \gc show significant iron spread, such as $\omega$ Cen \citep{marino2011}, possibly M22 \citep{marino2011b}, M54 \citep{carretta2010}, Terzan 5 \citep{origlia2011} and Liller 1 \citep{chiappino26} . Additionally, only a handful of metal-poor \gc show potential star-to-star dispersion in neutron-capture elements; for instance, \citet{Otsuki06} studied six red giant branch (RGB) stars in M15 and found significant variations in [Eu/Fe], [La/Fe], and [Ba/Fe]. According to the model presented by \citet{Otsuki06}, this spread in heavy elements could be reproduced if a weak r-process enriched supershell were uniformly followed by the main r-process.

NGC~4372 is one of the most metal-poor Galactic GC \citep[{[Fe/H]}$\approx$$-$2.2,][]{harris10, dias2016, sanroman}, with a present-day mass of $(1.89 \pm 0,07) \times 10^5 \rm M_{\sun}$ \citep[based on Gaia DR3 data and N-body models;][]{Baumgardt2021} and an age of 12.81 $\pm$ 0.81 Gyr \citep{Gontcharov2026}. 

Historically, the characterization of its stellar population began with the photometric studies carried out by \cite{hartwick1973} and \cite{alcaino1974}, who found a CMD characteristic of very metal-poor clusters, with large and variable absorption and a well-defined horizontal branch (HB) extending far to the blue.
Indeed, NGC~4372 suffers from differential reddening due to its location behind a strip of Galactic dust and gas; \citet{hartwick1973} suggested a reddening E(B-V) between 0.42 and 0.48 mag. This was confirmed in the photometric catalog of 39 Galactic \gc in the southern hemisphere \citep{rosenberg2000}, while \cite{kach} found a mean E(B-V)$\sim$0.5 mag with a significant variation between 0.3 and 0.8 mag across the cluster field of the GC. \citet{zinn} provided the first metallicity estimate for NGC 4372, obtaining a value of [Fe/H]=$-$2.08 $\pm$ 0.15. A decade later, \citet{geisler95} derived a mean metallicity of [Fe/H]=-2.10 $\pm$ 0.04 (excluding one metal-poor outlier) using the Ca \textsc{ii} triplet technique on a sample of 10 giant stars.
\citet{rut1997} estimated [Fe/H] values of $-$2.03 $\pm$ 0.03 on the \citet[][ZW84]{zinn} scale and $-$1.88 $\pm$ 0.05 on the \citet[][GC97]{gc97} scale.
More recently, \cite{carretta2009c} defined an updated metallicity scale using high S/N spectroscopic measurements of 19 \gc. Based on this scale, they provided a value of [Fe/H] = $-$2.19 $\pm$ 0.08 using the Ca \textsc{ii} data from \cite{geisler95}. Furthermore, \citet{dias2016} defined a new metallicity scale based on 51 Galactic GCs, finding a mean metallicity of [Fe/H]= $-$2.22 $\pm$ 0.06 for this cluster. Finally, \citet{massari2019} classified NGC~4372 as a Main Disc cluster based on its dynamical properties, pointing toward an \textit{in-situ} formation.

\citet[][hereafter SR15]{sanroman} presented the first detailed chemical abundance study of NGC~4372, based on UVES spectra of seven RGB stars acquired as part of the \textit{Gaia}-ESO Survey (GES). Their atmospheric parameters and chemical abundances were derived following the GES protocol, which included the standard linelist, MARCS models, solar abundances, and a synthetic spectrum library \citep[the entire methodology is described in the GES paper,][]{GES}. Abundances for Mg, Al, Si, Ca, Ti, Cr, and Ni were measured using the equivalent width (EW) method, while O, Na, Sc, Y, Ba, and La were determined through spectral synthesis. Although their original sample consisted of seven stars, only six were included in the final analysis; star $\#7$ was discarded due to its chemical peculiarities and discrepant metallicity.
They found a metallicity of [Fe/H] = $-$2.19 $\pm$ 0.03 with no evidence of metallicity spread. Additionally, they reported that NGC~4372 exhibits a Na-O anticorrelation and that its O abundances are relatively high compared to other galactic \gc. They also found a Mg-Al spread, suggesting that the Mg-Al burning cycle is active.

In the chemical composition space, \cite{horta2020} analyzed $\alpha$-elements abundances as a function of \feh, using the APOGEE DR16 data for 46 \gc. They focused on revealing how the kinematic classification of \citet{massari2019} correlates with the chemical composition space. For this purpose, they used Si abundances because they provide the most reliable abundance measures in APOGEE and do not show trends with stellar parameters. Following this approach, the chemical properties of NGC 4372 can be used to independently verify its dynamical classification as a Main Disc cluster. If we adopt the Si abundance from \cite{sanroman} ([Si/Fe] = 0.48 $\pm$ 0.02), the cluster aligns with the high-$\alpha$ plateau characteristic of the Main Disc population at low metallicities ([Fe/H] < $-$2.0). Although this value is slightly higher than the average reported for the APOGEE sample shown in \cite{horta2020}, it is consistent with the expected enrichment for an $\textit{in situ}$ system and lacks the $\alpha$-depletion observed in accreted GCs. This chemical signature provides robust independent evidence supporting the Main Disc origin of NGC 4372.

In this paper, we present a detailed chemical study of the metal-poor GC NGC~4372 based on new spectroscopic data. We derived abundances for 17 chemical elements, including light, $\alpha$, iron-peak, and heavy elements; notably, we provide the first measurements of Eu, a typical r-process element, for this cluster. A key advantage of our analysis is that stellar parameters were derived from the spectra; consequently, our results are independent of the uncertainties associated with interstellar reddening. By comparing these abundance distributions with field stars and other Galactic GCs, we aim to characterize the chemical evolution of NGC~4372 and determine whether it hosts MPs, providing new insights into its formation history.

In Section~\ref{obs}, we provide details regarding the observations and data reduction, while in Section~\ref{param}, we describe the derivation of stellar parameters for our sample. We present our results and analysis in Section~\ref{results}; finally, our conclusions are summarized in Section~\ref {conclusions}.

\section{Observations and data reduction}\label{obs}
\begin{figure}[h]
\centering
\includegraphics[width=\columnwidth]{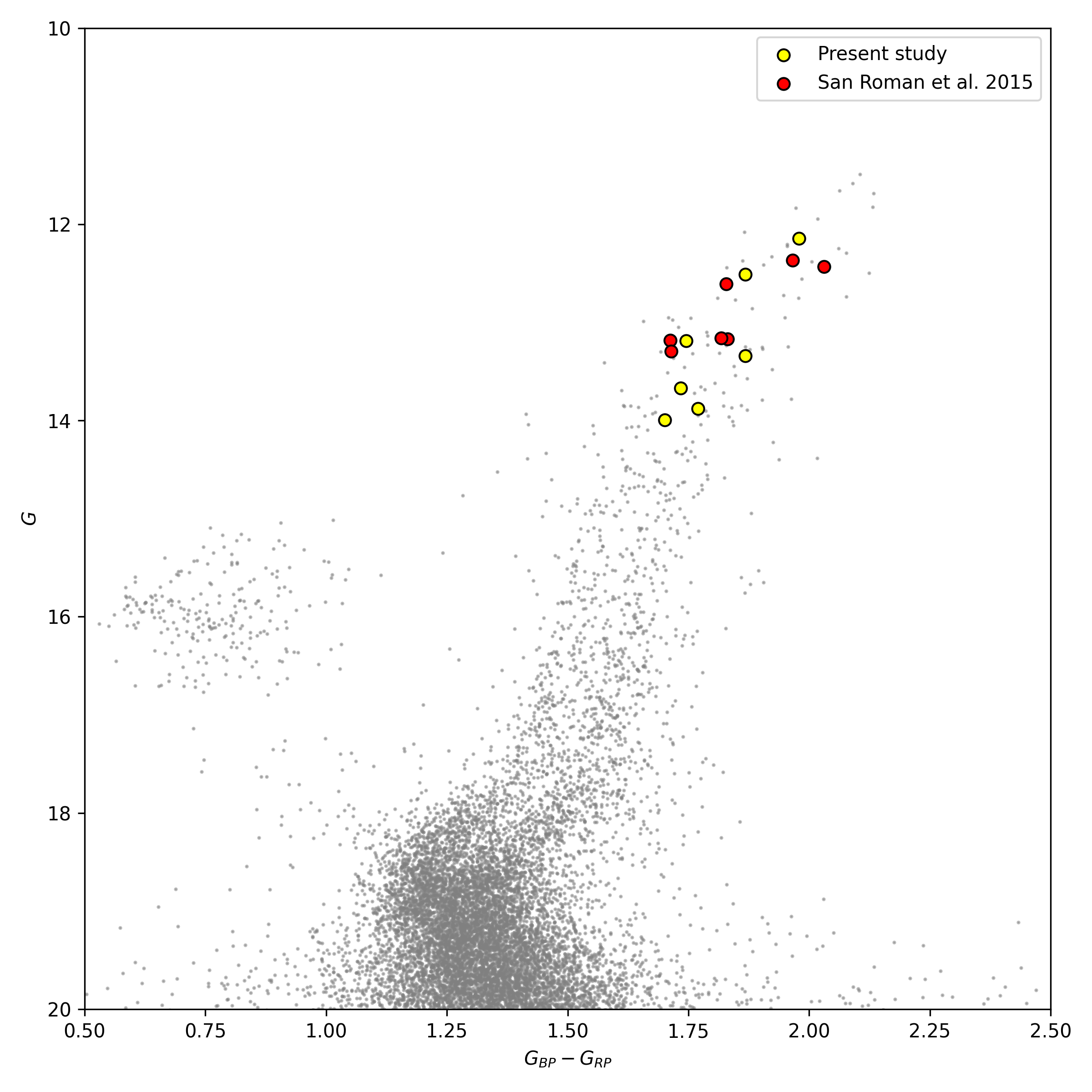}
 \caption{\textit{Gaia} color-magnitude diagram of NGC 4372. The background gray points represent the cluster catalog from \citet{vasiliev21}. The yellow circles highlight the seven RGB stars analyzed in the present study. For comparison, the red circles denote the RGB sample from \citetalias{sanroman}.}
 \label{fig:cmd}
\end{figure}

\begin{figure}
 \includegraphics[width=\columnwidth]{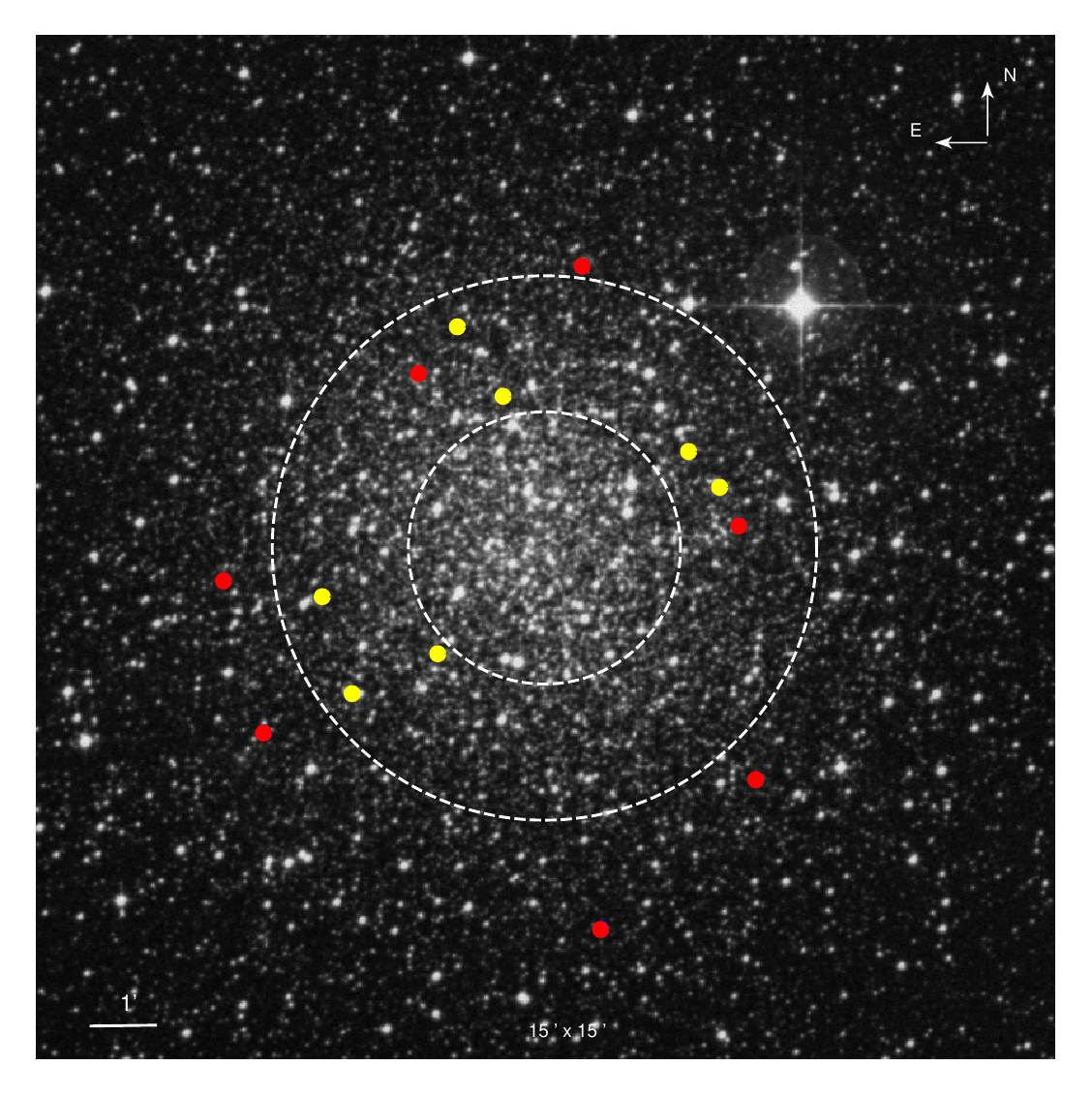}
 \caption{Spatial distributions of targets used in the present study (yellow symbols) and the sample of \citetalias[][red symbols]{sanroman}. Dashed circles correspond to the core radius, $r_{\rm c} = 1.7'$ \citep[][]{harris10}, and the half-light radius, $r_{\rm h} = 3.4'$ \citep{kach}} 
 \label{fig:spatial}
\end{figure}

Data were obtained in 2012 (ESO program $088.B-0492(A)$) using the FLAMES/UVES spectrograph mounted on the ESO VLT/UT2 telescope at Cerro Paranal, Chile. The UVES spectrograph provides a spectral resolution of $R \simeq 47000$. Observations were conducted with a central wavelength at $\sim 580$ nm, covering a range from $480-670$ nm, with an exposure time of 2775 s per observation. To increase the signal-to-noise ratio (S/N), we stacked three spectra for each target, achieving an S/N between 40 and 60 at 650 nm. Our sample includes seven Red Giant Branch (RGB) stars, as shown in the cluster CMD (Fig.~\ref{fig:cmd}). While \citetalias{sanroman} also analyzed seven RGB stars from GES using the same instrument, there are no stars in common between our samples, as illustrated in Fig.~\ref{fig:spatial}. In this figure, our spectroscopic targets are indicated as yellow symbols, whereas the \citetalias{sanroman} sample targets are represented by red symbols. The spatial distribution clearly reveals that there is no spatial overlap between the two datasets across the cluster field.

The dataset was reduced using the ESO CPL UVES pipelines $5.7.0$\footnote{\url{http://eso.org/sci/software/pipelines/}}, which included bias subtraction, flat-field correction, and wavelength calibration. Subsequently, we used \textsc{iraf}\footnote{\textsc{iraf} is distributed by the National Optical Astronomy Observatory, which is operated by the Association of Universities for Research in Astronomy, Inc., under a cooperative agreement with the National Science Foundation.} to perform sky subtraction using the \textit{sarith} task and spectral normalization with the \textit{continuum} task. Following this, we co-added the spectra for each star using the \textit{scombine} task. Radial velocities (RVs) were measured via cross-correlation with a synthetic spectrum of similar stellar parameters using the \textit{fxcor} task. Finally, each RV was corrected to the heliocentric system with the \textit{rvcorrect} task. 
Through this procedure, we obtained a mean RV of $79.56 \pm 1.0$ \km with a dispersion of $2.38$ \km. In comparison \citet{Baumgardt2021} derived a RV of $75.59$\km with $1\sigma$ error of 0.30 km s$^{-1}$ using \textit{Gaia} DR3 data. Furthermore, \citet{lardo2015}, as part of the GES, used a sample of over 100 stars with FLAMES/GIRAFFE and obtained a mean RV of $75.2 \pm 0.4 $ \km with a dispersion of 3.9 \km. \cite{kach} presented a kinematic study based on FLAMES-GIRAFFE observations of 131 cluster members, reporting a heliocentric RV of $75.9 \pm 0.4$ \km with a dispersion of 3.8 \km (see Table \ref{tab:stellar}).
Other measurements include those \citetalias{sanroman}, with a mean of $72.6 \pm 1.3$ \km and a dispersion of 3.6 \km, and  \cite{geisler95}, with a mean of $73.2 \pm 1.4$ \km, (where the error represents the standard error of the mean). Our mean heliocentric RV is $\sim$5 \km higher than the values reported by \citet{lardo2015} and \citet{kach}, which is likely attributable to our smaller sample size and differences in data processing.

\begin{figure}[h]
\includegraphics[width=\columnwidth]{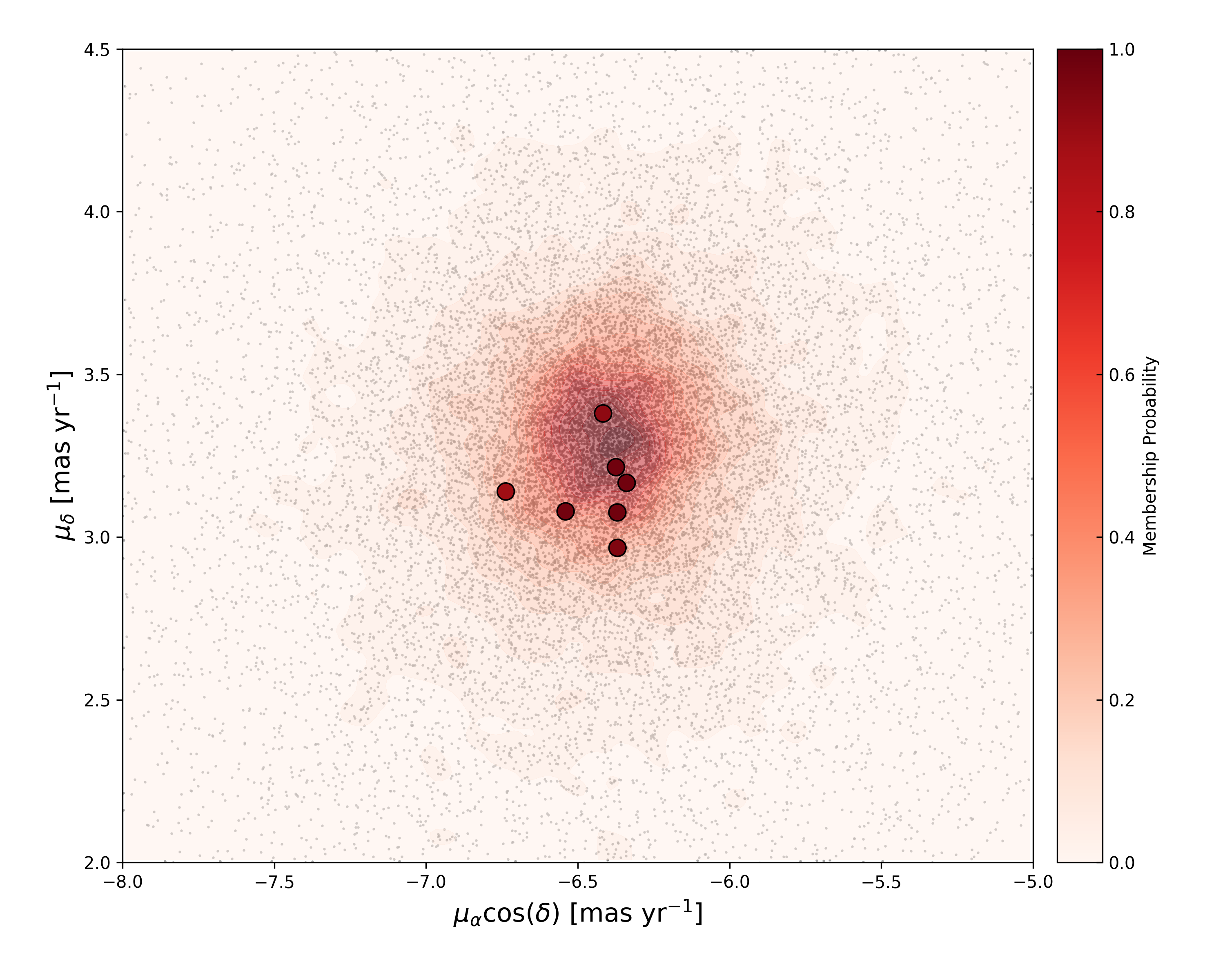} 
\caption{The Gaia DR3 proper motion distribution of stars of the GC NGC~4372. The background density plot and the side color bar indicate the membership probability of individual stars, according to the catalog by \citet{vasiliev21}. Red filled circles represent our seven UVES targets for this study, all of which are located in the high-probability region (membership probability $\approx$ 1).}
\label{fig:gaia_pm}
\end{figure}

In Fig.~\ref{fig:gaia_pm}, we present the proper motion (PM) from the Gaia DR3 catalog for stars within 15 arcmin from the cluster$^\prime$s center (\cite{brown18}; \cite{lindergren18}); the red filled circles are our sample of seven stars. We note that the position of our targets is consistent with the PM distribution of NGC 4372. Furthermore, the radial velocity of our sample (Table~\ref{tab:stellar}), the positions in the Gaia broad-band CMD (Fig.~\ref{fig:cmd}), and the chemical patterns discussed in Section~\ref {param} collectively confirm the cluster membership of our objects.

Table~\ref{tab:stellar} summarizes the basic properties of the observed stars, including their Gaia DR3 IDs, coordinates (RA and DEC in degrees), signal-to-noise (S/N), $G-$band magnitudes, BP$-$RP colors, radial velocities, and proper motions ($\mu_{\alpha}$, $\mu_{\delta}$). Additionally, the table lists the adopted atmospheric parameters, specifically effective temperatures (\teff) and surface gravities (\logg), along with the derived metallicities (\feh).

\begin{table*}[h!]
 \centering
 \caption{Stellar parameters and observational data for the selected sample of the seven observed stars in NGC 4372.}
 \label{tab:stellar}
 \resizebox{\textwidth}{!}{%
 \begin{tabular}{cccccccccccccc}
 \hline
 ID & RA & DEC & SNR & G & BP$-$RP & $v_{\rm rad}$ & $\mu_{\alpha}$ & $\mu_{\delta}$ & \teff & \logg & [Fe/H]\\
 & [$^\circ$] & [$^\circ$] & & [mag] & [mag] & \km & [mas/year] & [mas/year] & [K] & [dex] & \\
 \hline
 5841718339382688512 & 186.604968 & $-$72.691746 & 50.62 & 13.1867 & 1.7456 & 77.8 & $-$6.5402 & 3.0789 & 4434 & 0.67 & $-$2.39\\
 5841722599990467712 & 186.464804 & $-$72.621452 & 40.80 & 13.8789 & 1.7698 & 77.0 & $-$6.7372 & 3.1400 & 4605 & 0.90 & $-$2.41\\
 5841721259960993664 & 186.532350 & $-$72.683354 & 43.19 & 13.6707 & 1.7341 & 80.5 & $-$6.339 & 3.1665 & 4500 & 0.90 & $-$2.34\\
 5841722698761878784 & 186.498556 & $-$72.603745 & 43.92 & 13.3433 & 1.8681 & 79.0 & $-$6.3742 & 3.2147 & 4480 & 0.86 & $-$2.35\\
 5841722149006071552 & 186.292204 & $-$72.647087&  56.85 &  12.1451 &  1.9788 & 84.4 & $-$6.4166 & 3.3799 &  4326 & 0.55 & $-$2.19\\
 5841721393091831168 & 186.624055 & $-$72.667572 & 57.62 &  12.5095 &  1.8679 & 79.0 & $-$6.3692 & 2.9667 & 4375 & 0.56 & $-$2.36\\
 5842472879233356160 & 186.315798 & $-$72.637815 & 43.09 & 13.9942 & 1.7007 & 79.2 & $-$6.3697 & 3.0756 & 4662 & 0.54 & $-$2.45\\
 \hline
 \end{tabular}%
 }
 \tablefoot{The columns include the Gaia DR3 identifier, equatorial coordinates (RA and DEC in degrees), proper motions ($\mu_{\alpha}$, $\mu_{\delta}$), G$-$band magnitudes, and BP-RP colors. The table also lists the radial velocities and the adopted atmospheric parameters: effective temperatures (\teff), surface gravities (\logg) and metallicity (\feh).}
\end{table*}

\section{Atmospheric parameters and abundance analysis}\label{param}
The Local Thermodynamic Equilibrium (LTE) program \textsc{moog} \citep{sneden73} \footnote{Freely distributed by C. Sneden, University of Texas at Austin} was used to determine chemical abundances. \textsc{moog} is an LTE line analysis and synthetic spectrum code that solves the radiative transfer equations to derive abundances, either by matching the EWs of atomic lines or through a best-fit synthetic spectrum approach, particularly for blended features and molecules. The linelist for our analysis was compiled from various literature sources, following the compilation by \citet[see][]{villa2011}. To ensure high-precision results, we performed a solar-inverse calibration using the spectral synthesis method described in \cite{villanova09}. In this approach, the oscillator strengths $\log gf$ values of the selected lines were empirically adjusted so that the synthetic solar spectrum, generated with a standard solar model atmosphere, reproduced the reference solar abundances, that is listed in Table~\ref{tab:abun}. This calibration removes potential systematic offsets and minimizes the scatter in abundances relative to the mean. This specific linelist and methodology have been extensively validated in previous studies \citep{Rain_2019,rup106,munoz17,munoz18,munoz20,mura18}.

Atmospheric models for each star were calculated using interpolated grids of  \textsc{atlas9} model atmospheres \citep{kuru}. The initial model for each star was constructed using a first guess of \teff, \logg, \vt, and the [Fe/H] value from \cite{harris10}. Specifically, we adopted the stellar parameters of the representative star ($\#3$) of \citetalias{sanroman} sample as the starting point for our analysis. To refine these parameters, we performed an iterative spectroscopic analysis. The effective temperature \teff was determined by enforcing excitation equilibrium of Fe I lines, requiring a null correlation between the derived line-by-line abundances and their respective Excitation Potential (E.P). The larger number and range of Fe I lines ensure a more precise determination of \teff and minimize the impact of individual line uncertainties. The microturbulence velocity \vt was adjusted to eliminate any trend between Fe I abundances and the reduced EW. Once the \teff and \vt were determined, we adjusted the value of \logg to satisfy the ionization equilibrium (i.e., abundances derived from Fe I and Fe II must be nearly identical, with a maximum difference of 0.02 dex). Finally, the metallicity [Fe/H] was derived from the iron abundance relative to the Sun, as determined by the abundance analysis. Since the parameters are interdependent, the procedure was iterated until a converged, self-consistent solution was reached.

To validate our spectroscopic temperatures, we compared them with the values derived from the color-temperature relations of \citet{mucciarelli20}. We found an average difference of 260 K compared to our determinations (see Table \ref{tab:comp}). This systematic offset is likely due to the inherent limitations of LTE assumptions in the atmospheric models of metal-poor giants, where non-LTE (NLTE) effects can be significant. Nevertheless, we have retained our spectroscopically derived parameters (Table \ref{tab:stellar}) to ensure methodological consistency with previous studies of this and similar GCs \citep{sanroman,Rain_2019,roederer2015,hanke2017}, thereby facilitating a direct comparison of the chemical evolution trends.

NGC~4372 suffers from significant differential reddening due to its location behind a dense strip of Galactic gas and dust. While the Harris catalogue provides a mean color excess of E(B-V)$=0.39$ \citep{harris10}, \cite{kach} found a higher mean value of E(B-V) $\sim 0.5$, with substantial variations ranging from 0.3 to 0.8 mag across the field. In our analysis, because the stellar parameters were derived iteratively and directly from the spectra, they remain unaffected by the uncertainties and variations in interstellar reddening.

\begin{figure} 
 \includegraphics[width=\columnwidth]{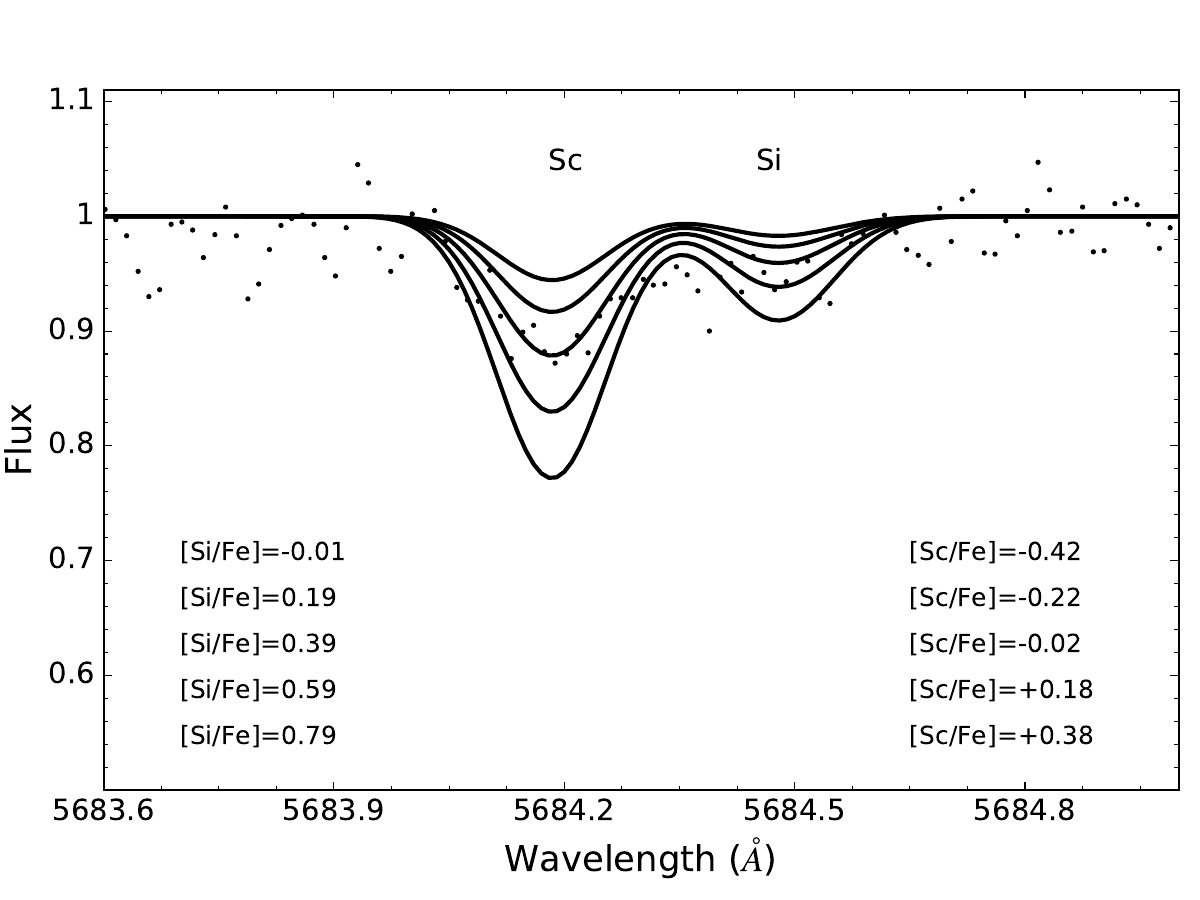}
 \caption{Fitting process in star $\#3$ for Sc and Si absorption lines. Continuous lines are synthetic spectra, while points correspond to the observed spectrum.}
 \label{fig:ss}
\end{figure}

Chemical abundances for Fe, Ca, Ti, Cr, and Ni were derived from the EWs of their respective spectral lines (typically ranging from 5 to 40 lines, depending on the species). For the remaining elements (Na, O, Mg, Al, Si, Sc, Zn, Y, Ba, La, Ce, Eu, and Nd), where lines are often affected by blending or complex profiles, we employed the spectrum-synthesis method of \textsc{moog}. For each feature, we generated a grid of five synthetic spectra with varying abundances, centered around an initial estimate. The code solves the radiative transfer equation under LTE conditions to produce a theoretical profile that accounts for instrumental, thermal, and rotational broadening. The final abundance was determined by identifying the synthetic profile that minimized the root-mean-square (rms) residuals between the observed and calculated spectra.
In Fig.~\ref{fig:ss} we show an example of the fitting process used to derive Sc and Si abundances for star $\#3$. To ensure the reliability of the fits, any spectral regions contaminated by telluric absorption or cosmic rays were strictly excluded from the analysis.  

The linelist for the chemical analysis has been described in previous studies \citep[e.g][]{villa2011,line2013}: Sodium (Na) was measured using the lines at 5682-5688 \r{A}, and the resulting abundances were corrected for NLTE using the corrections provided by the INSPEC \footnote{\url{http://inspect.coolstars19.com/index.php?n=Main.HomePage}} database \citep{Lind11}. Oxygen (O) abundances are based on the forbidden [OI] line at 6300 \r{A}. Mg abundances are typically derived from the three high-excitation lines: 5711, 6318, and 6319 \r{A}. However, in this study, it was obtained from the line at $~$5711.09 \r{A}, while Al was obtained from the line at 6696 \r{A}. Si and Sc abundances were obtained from the lines at 5684 \r{A} (Fig.~\ref{fig:ss}), while Zn was obtained from the 4811 \r{A} line. Finally, we determined abundances of several heavy elements: La (5123 \r{A}), Y (4900 \r{A}), Ba (5853 \r{A}), Ce (5274 \r{A}), Eu (6645 \r{A}) and Nd (5320 \r{A}). For Ba, hyperfine splitting (HFS) was taken into account due to its strength and profile complexity. The complete set of measured elements and their respective abundance ratios for each star are presented in Table ~\ref{tab:abun}. For some stars, reliable measurements for Al, La, Ce, Eu, and Nd could not be obtained because the lines are too weak.

For an internal error analysis of the atmospheric parameters, we selected star \#3 as representative of our sample. We do not scale the resulting abundance uncertainties with the signal-to-noise ratio (S/N) for the other stars because the S/N is highly consistent across our entire sample. Furthermore, the contribution of the S/N to the total abundance error budget is minimal compared to the uncertainties introduced by the atmospheric parameters themselves, meaning that any S/N-based scaling would have a negligible impact on the final reported errors. We followed the uncertainty estimation procedure described by \citet{marino2008}, where the internal uncertainties in effective temperature ($\Delta$\teff), surface gravity ($\Delta$\logg), microturbulent velocity ($\Delta$\vt), and metallicity ($\Delta$[Fe/H]) are derived directly from the spectroscopic data. $\Delta$\teff was obtained by varying the temperature until the slope of the line that best fits the relation between Fe I abundances and E.P. became equal to the error on the fit. $\Delta$\vt was determined by varying the microturbulence until the slope of the line that best fits the relation between Fe I abundances and reduced EW became equal to the error on the fit. $\Delta$\logg was set by requiring that the following relation is satisfied:

\begin{equation}
    [\rm Fe I/H]-\sigma_{[FeI/H]} = [FeII/H]+\sigma_{[FeII/H]}
\end{equation}

and $\Delta$[Fe/H] corresponds to the statistical uncertainty in the mean iron abundance. The resulting parameter uncertainties ($\Delta$\teff $=35$ K, $\Delta$\logg $= 0.12$, $\Delta$[Fe/H]$=0.03$ dex, and $\Delta$\vt $=0.04$ \km) thus represent the measured sensitivities and internal precision of our spectroscopic criteria.

After that, we redetermined abundances for star $\#3$. We also performed a sensitive analysis using more conservative parameter uncertainties: $\Delta$\teff $=70$ K, $\Delta$\logg $= 0.24$,  $\Delta$[Fe/H]$=0.06$ dex, and $\Delta$\vt $=0.08$ \km. Table~\ref{tab:errors} (cols. 2-5) lists the resulting abundance errors for each atmospheric parameter, with values corresponding to these more conservative assumptions in parentheses.

The uncertainty associated with the spectral noise ($\sigma_{\rm S/N}$) was calculated according to the measurement method employed. For elements measured via EWs, $\sigma_{\rm S/N}$ was estimated by dividing the rms scatter of the derived abundances by the square root of the number of lines used. For those determined through spectral synthesis, we adopted the formal error provided by the fitting procedure, which accounts for the residuals between the observed and synthetic profiles. The total internal error ($\sigma_{\rm tot}$) for each abundance ratio [X/Fe] was then determined by the quadratic sum of the propagated uncertainties from atmospheric parameters and $\sigma_{\rm S/N}$:
\begin{equation}
    \sigma_{\text{tot}} = \sqrt{ \sigma_{\text{S/N}}^2 + \sigma_{T_{\text{eff}}}^2 + \sigma_{\log g}^2 + \sigma_{v_t}^2 + \sigma_{[\text{Fe/H}]}^2 }
\end{equation}

Finally, the total observed dispersion for the entire sample ($\sigma_{\text{obs}}$), which encapsulates both individual measurement errors and any potential intrinsic astrophysical variation, is computed as:
\begin{equation}
    \sigma_{\text{obs}} = \sqrt{\frac{\sum_{i=1}^{N} (X_i - \bar{X})^2}{N - 1}}
\end{equation}
where $N$ is the number of target stars, $X_i$ is the abundance of the individual star, and $\bar{X}$ is the sample mean (as presented in Table~\ref{tab:errors}).

\begin{table*}
\centering
\begin{threeparttable}
 \caption{Chemical abundances of the observed stars.}
 \label{tab:abun}
 \begin{tabular}{lccccccccc}
 \hline
  El. & $\#$1 & $\#$2 & $\#$3 & $\#$4 & $\#$5 & $\#$6 & $\#$7 & NGC~4372 & Sun\tnote{*} \\
  \hline
  {[Na/Fe]}$_{\rm NLTE}$ & 0.58 & 0.36 & 0.54 & 0.33 & 0.74 & 0.16 & 0.69 & 0.49 $\pm$ 0.08 & 6.32 \\
  {[O/Fe]} & 0.23 & 0.45 & 0.25 & 0.53 & 0.11 & 0.43 & 0.33 & 0.33 $\pm$ 0.06 & 8.83\\
  {[Mg/Fe]} & 0.42 & 0.19 & 0.34 & 0.47 & 0.27 & 0.38 & 0.38 & 0.35 $\pm$ 0.04 & 7.56\\
  {[Al/Fe]} & 1.13 & 0.94 & 0.92 & - & 1.20 & - & 0.89 & 1.02 $\pm$ 0.06& 6.43\\
  {[Si/Fe]} & 0.35 & 0.19 & 0.42 & 0.58 & 0.22 & 0.52 & 0.55 & 0.40 $\pm$ 0.06 &7.61 \\
  {[Ca/Fe]} & 0.37 & 0.38 & 0.37 & 0.34 & 0.29 & 0.35 & 0.41 & 0.36 $\pm$ 0.02 & 6.39\\
  {[Ti/Fe]} & 0.24 & 0.22 & 0.21 & 0.26 & 0.21 & 0.14 & 0.25 &0.22 $\pm$ 0.02  & 4.94\\
  {[Sc/Fe]} & -0.06 & -0.23 & -0.04 & -0.05 & -0.23 & -0.06 & -0.08 & -0.11 $\pm$ 0.03& 3.12\\
  {[Cr/Fe]} & -0.31 & -0.34 & -0.28 & -0.29 & -0.24 & -0.32 & -0.21 & -0.29 $\pm$ 0.02 & 5.63\\
  {[Fe/H]} & -2.39 & -2.41 & -2.34 & -2.35 & -2.19 & -2.36 & -2.45 & -2.36$\pm$0.03 & 7.50 \\
  {[Ni/Fe]} & -0.07 & 0.17 & -0.07 & -0.11 & -0.12 & -0.16 & -0.06 & -0.06 $\pm$ 0.04& 6.26\\
  {[Zn/Fe]} & 0.08 & 0.34 & 0.25 & -0.03 & -0.01 & -0.01 & 0.14 & 0.11 $\pm$ 0.05 & 4.61\\
  {[Y/Fe]} & -0.17 & -0.40 & -0.37 & -0.15 & -0.06 & -0.25 & -0.36 & -0.25 $\pm$ 0.05 & 2.25\\
  {[Ba/Fe]} & -0.47 & -0.57 & -0.51 & -0.39 & -0.30 & -0.57 & -0.51 & -0.47 $\pm$ 0.04 & 2.34\\
  {[La/Fe]} & 0.03 & - & 0.15 & 0.18 & -0.04 & -0.10 & 0.32 & 0.09 $\pm$ 0.06 & 1.26\\
  {[Ce/Fe]} & 0.04 & - & -0.16 & - & -0.29 & -0.20 & -0.26 & -0.17 $\pm$ 0.06 & 1.53\\
{[Eu/Fe]} & 0.24 & 0.44 & 0.28 & 0.62 & 0.46 & 0.36 & - & 0.40 $\pm$ 0.06& 0.52\\

  {[Nd/Fe]} & 0.06 & -0.18 & 0.07 & 0.21 & 0.11 & -0.11 & - & 0.03 $\pm$ 0.06& 1.59\\
  \hline
 \end{tabular}
\end{threeparttable}
\tablefoot{Columns 2-8: Chemical abundances of the observed stars. Column 9: Mean abundance of NGC 4372 and its statistical error. Column 10: Solar abundance used for the analysis (indicated as $\log_{\epsilon}$(El)). [Ti/Fe] is the average between TiI and TiII. Solar abundances from \cite{villa2011}}

\end{table*}

\begin{table}
 \centering
 \caption{Determination of effective temperature.}
 \label{tab:comp}
 \begin{tabular}{ccc}
 \hline
 Teff$_{Bel}$[K] & Teff$_{Mu}$[K] & $\Delta$[K]  \\
 {\tiny This study} & {\tiny Mucciarelli et al. 2020} &  $|$Teff$_{Bel} -$  Teff$_{Mu}|$ \\

 \hline
 4434 & 4707 &273 \\
 4605 & 4883 & 278 \\
 4500 & 4760 & 260\\
 4480 & 4742 & 262\\
 4326 &4546& 220\\
 4375 & 4640 &265 \\
 4662 & 4951 &  289\\
 \hline
 \end{tabular}
 \tablefoot{Determination and comparison of effective temperature using \citet{mucciarelli20} relation.}
\end{table}

\begin{table*}
 \centering
 \caption{Estimated errors on abundances.}
 \label{tab:errors}
 \begin{tabular}{lccccccr}
 \hline
  ID & $\Delta$\teff =35(70) K & $\Delta$\logg =0.12(0.24) & $\Delta$\vt =0.04(0.08) & $\Delta$\feh =0.03(0.06) & $\sigma_{\rm S/N}$ & $\sigma_{\rm tot}$ & $\sigma_{\rm obs}$ \\
  \hline
  $\Delta$([Na/Fe])& 0.01(0.02) & 0.01(0.03) & 0.02(0.06) & 0.06(0.12) & 0.02 & 0.07(0.14) &0.20\\
  $\Delta$([O/Fe]) & 0.02(0.04)  & 0.02(0.06)  & 0.00(0.02)  & 0.01(0.02) & 0.02 & 0.04(0.08) &0.14\\
  $\Delta$([Mg/Fe])& 0.02(0.04) & 0.01(0.02) & 0.01(0.02) & 0.02(0.03) & 0.02 & 0.04(0.06) &0.09\\  
  $\Delta$([Al/Fe])& 0.02(0.09) & 0.02(0.06) & 0.01(0.08) & 0.02(0.10) & 0.02 & 0.04(0.16) &0.14\\
  $\Delta$([Si/Fe])& 0.04(0.06) & 0.02(0.04) & 0.02(0.03) & 0.05(0.07) & 0.02 & 0.07(0.10) &0.16\\
  $\Delta$([Ca/Fe])& 0.03(0.03) & 0.01(0.03) & 0.00(0.03) & 0.01(0.03) & 0.02 & 0.04(0.06) & 0.04\\
  $\Delta$([Ti/Fe])& 0.05(0.06) & 0.01(0.06) & 0.01(0.06) & 0.00(0.06) & 0.05 & 0.07(0.13) & 0.04\\
  $\Delta$([Sc/Fe])& 0.01(0.06) & 0.03(0.10) & 0.02(0.02) & 0.01(0.08) & 0.03 & 0.05(0.14) &0.08\\  
  $\Delta$([Cr/Fe])& 0.017(0.03) & 0.02(0.02) & 0.01(0.02) & 0.01(0.02) & 0.04 & 0.05(0.06) & 0.05\\
  $\Delta$([Fe/H])& 0.015(0.02) & 0.02(0.02) & 0.01(0.02) & 0.01(0.02) & 0.02 & 0.04(0.04) & 0.03 \\
  $\Delta$([Ni/Fe])& 0.05(0.07) & 0.01(0.07) & 0.00(0.07) & 0.01(0.07) & 0.09 & 0.10(0.17) & 0.11\\
  $\Delta$([Zn/Fe])& 0.01(0.03) & 0.02(0.04) & 0.02(0.04) & 0.01(0.06) & 0.04 & 0.05(0.09) &0.14\\
  $\Delta$([Y/Fe]) & 0.01(0.07) & 0.01(0.10) & 0.01(0.03) & 0.02(0.09) & 0.04 & 0.05(0.15) &0.13\\
  $\Delta$([Ba/Fe])& 0.00(0.06) & 0.02(0.06) & 0.00(0.06) & 0.03(0.04) & 0.01 & 0.04(0.11) &0.10\\
  $\Delta$([La/Fe])& 0.01(0.02) & 0.02(0.03) & 0.01(0.06) & 0.01(0.03) & 0.03 & 0.04(0.08) &0.16\\
  $\Delta$([Ce/Fe])& 0.01(0.09) & 0.03(0.04) & 0.00(0.08) & - & 0.03  &0.13(0.04) & 0.13\\
  $\Delta$([Eu/Fe])& 0.02(0.10) & 0.04(0.06) & 0.00(0.03) & 0.00(0.04) & 0.02 & 0.05(0.12) &0.14\\
  $\Delta$([Nd/Fe])& 0.02(0.05) & 0.03(0.06) & 0.00(0.04) & 0.00(0.02) & 0.03 & 0.05(0.10) &0.15\\
  \hline
  \end{tabular}
  \tablefoot{Estimated errors on abundances due to errors on atmospheric parameters and to spectral noise, compared with the observed errors ($\sigma_{\rm obs}$). Values in parentheses are those calculated using the more conservative errors on the parameters.}
\end{table*}

\section{Results}\label{results}
In this work, we examined in detail the chemical patterns in NGC~4372 to search for the presence of MPs. Additionally, we compared our results with those obtained by other paper that studied the same cluster, as well as with field stars and other Galactic \gc of similar metallicity. In the following subsections, we present the chemical abundances for iron-peak, $\alpha$-, light, and heavy elements.

\subsection{Iron-peak elements}\label{ironpeak}

We obtained a mean metallicity value of \feh =$-$2.36 $\pm$ 0.03 dex, where the quoted uncertainty represents the standard error of the mean. The observed iron dispersion (see Table~\ref{tab:errors}) is consistent with our measurement uncertainties, suggesting no evidence of an intrinsic Fe abundance spread. Notably, while star $\#$5 is 0.16 dex more metal-rich than the sample average, it exhibits no other chemical or physical peculiarities that would justify its exclusion.

Our value is lower than those given by 
\citet[\feh=$-$2.08 $\pm$ 0.15]{zinn} and \citet[\feh=-2.10 $\pm$ 0.04]{geisler95}. However, these values were derived using a methodology that differs substantially from the one adopted in this paper.
On the other hand, our metallicity compares better with \citet[\feh=-2.19$\pm$ 0.08]{carretta2009c}, \citet[\feh= $-$2.22 $\pm$ 0.06]{dias2016} and \citet[\feh=$-$2.19 $\pm$ 0.03]{sanroman}, where the authors used a methodology similar to that adopted here.
Systematic offsets of $\sim 0.15$ dex can still appear \citep{hinkel14} when comparing our results with certain recent literature estimates due to differences in model atmospheres, adopted line lists, atomic data, temperature determination methods, and solar reference abundances rather than inconsistencies in the analysis \citep{jofre19}.

The chemical abundances for the iron-peak elements Sc, Cr, Ni, and Zn are listed in Table~\ref{tab:abun}. In general, our sample has abundances of these elements that coincide well with those of other GCs and halo field stars of similar metallicity, and we found no intrinsic dispersion among the sample stars.

Scandium (Sc) follows the solar scale, with only two stars appearing slightly underabundant. This is consistent with the hypothesis that Sc remains solar-scaled across a wide range of metallicities \citep{prochaska2000}. Cr is subsolar, in good agreement with \citetalias{sanroman}, and follows the typical trend of similarly \gc and halo field stars, increasing with metallicity. Similarly, Ni is essentially solar-scaled; this is a common feature across the entire metallicity range, indicating that Ni tracks Fe production and suggesting their nucleosynthetic origins are closely linked. This study also provides the first Zn abundance measurement for this cluster, which we found to be slightly overabundant ($\langle$[Zn/Fe]$\rangle=$0.11). While
\citet{duffau2017} analysed a large sample of 1301 stars from the GES iDR4 data and found that stars at low metallicities (\feh $<$ 0.5) show slightly supersolar [Zn/Fe] and constant at all galactocentric radii, their sample only extended down to [Fe/H]$\sim$-1.1. In a more metal-poor regime, \citet{nissen2007} showed that halo stars between -2.7<\feh<-2 have, on average [Zn/Fe]$\sim$ 0.1 with a sharp increase to $\sim$ 0.5 at metallicities below \feh = -3, and have values close to zero in the metallicity range of -2<\feh<-1. Our result is fully compatible with the galactic trend.

\subsection{\texorpdfstring{$\alpha$\xspace}-elements}\label{alpha-el}

\begin{figure}
 \includegraphics[width=\columnwidth]{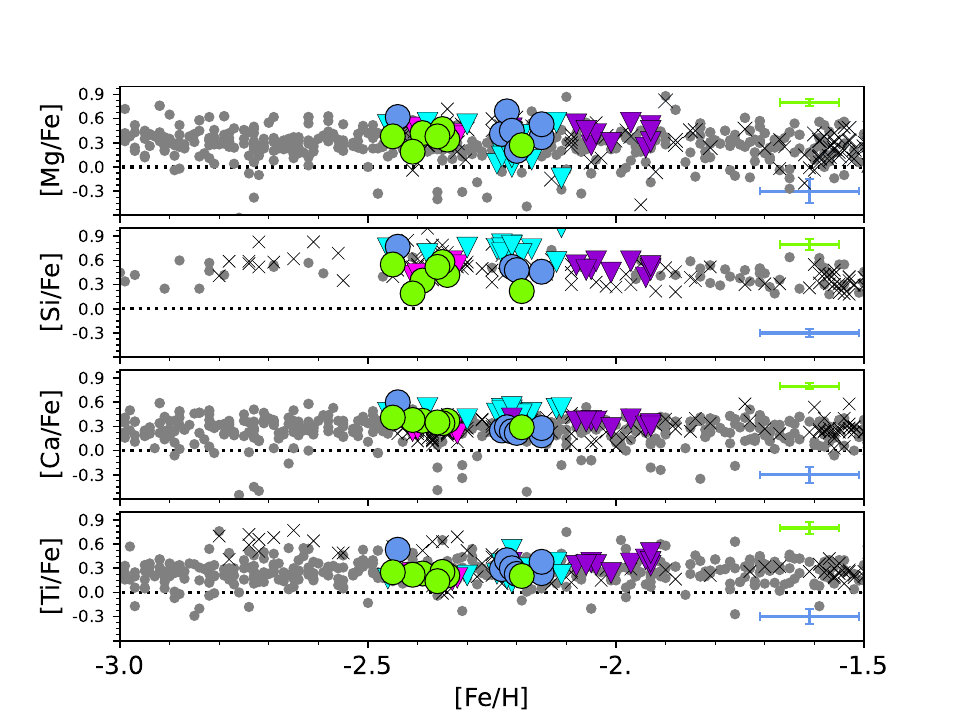}
 \caption{[Mg/Fe], [Si/Fe], [Ca/Fe], and [Ti/Fe] trends as a function of [Fe/H] for different environments. Green filled circles indicate our targets, light blue filled circles are data from \citetalias{sanroman}, black symbols are different \gc, and gray points are field stars. We also include abundances of metal-poor \gc: NGC~4833 \citep[cyan triangles,][]{roederer2015}, NGC~6426 \citep[magenta triangles,][]{hanke2017}, NGC~6809 \citep[violet triangles,][]{Rain_2019}.}
 \label{fig:alpha}
\end{figure}

\begin{figure}
 \includegraphics[width=\columnwidth]{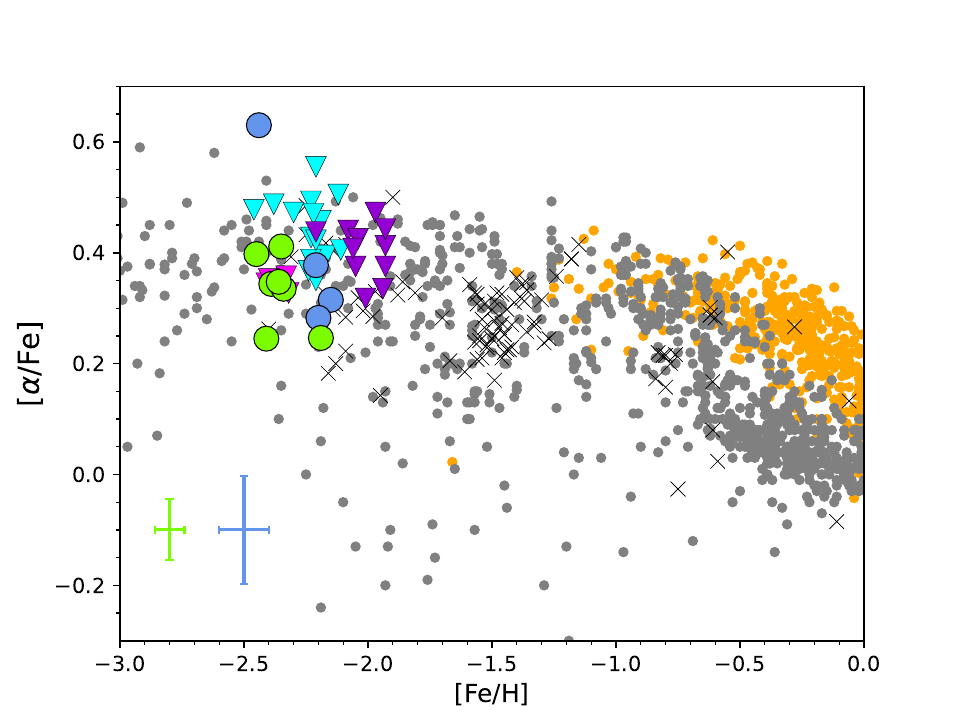}
 \caption{[$\alpha$/Fe] versus [Fe/H] for different environments. Green filled circles indicate our targets, light blue filled circles are data from \citetalias{sanroman}, black symbols are different \gc, gray points are field stars, and orange points are bulge field stars \citep{gonzalez2011}. We also include abundances of metal-poor \gc: NGC~4833 \citep[cyan triangles,][]{roederer2015}, NGC~6426 \citep[magenta triangles,][]{hanke2017}, NGC~6809 \citep[violet triangles,][]{Rain_2019}.}
 \label{fig:alphaprom}
\end{figure}

\begin{figure}
 \centering
 \includegraphics[width=\columnwidth]{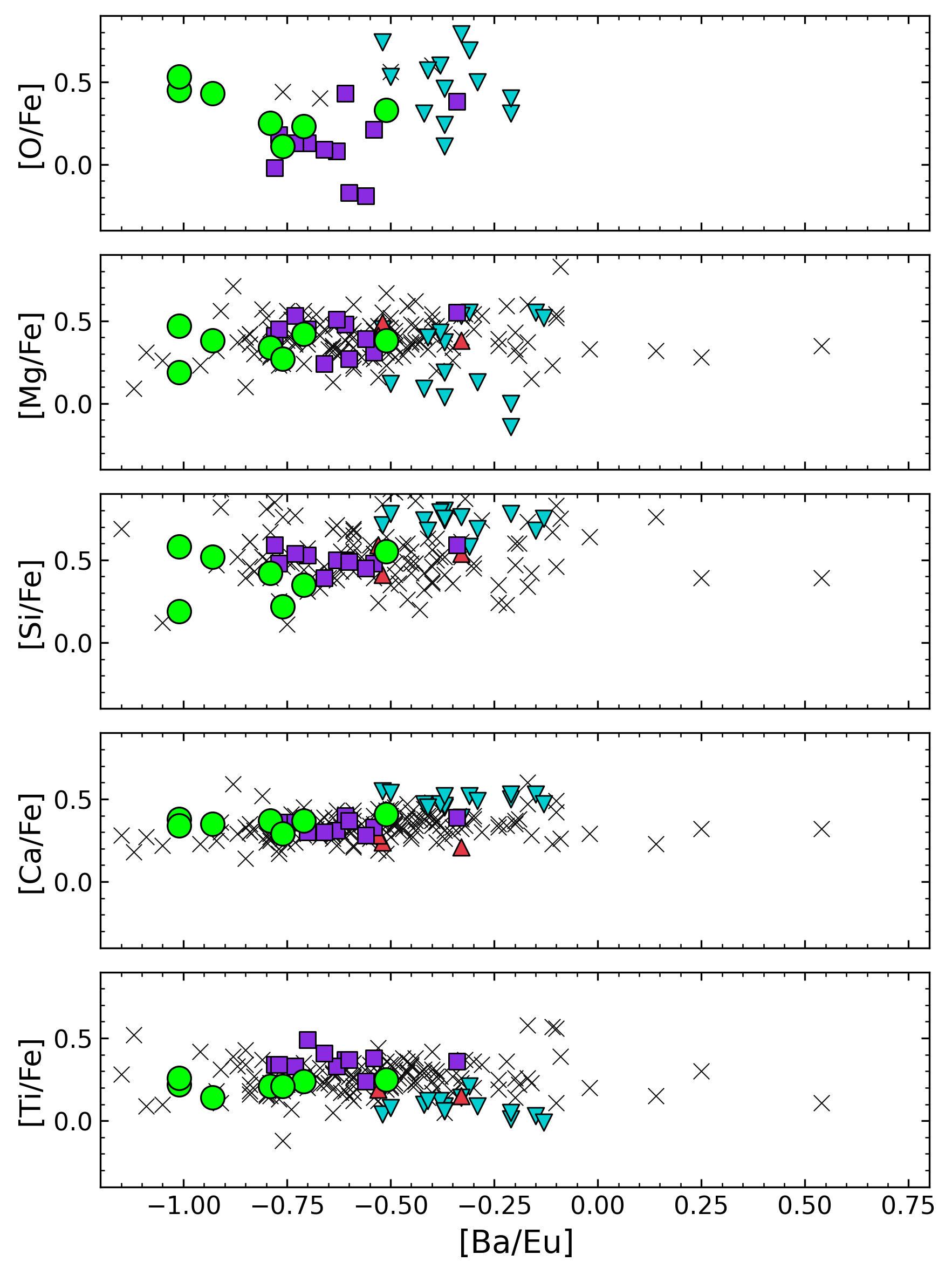}
 \caption{$\alpha$-element abundance ratios ($[\text{O/Fe}]$, $[\text{Mg/Fe}]$, $[\text{Si/Fe}]$, $[\text{Ca/Fe}]$, and $[\text{Ti/Fe}]$) as a function of $[\text{Ba}/\text{Eu}]$ for our targets in NGC~4372 (green circles). For comparison, Galactic halo field stars from \citet{sakari2018} (gray crosses) and other globular clusters are included: NGC~4833 \citep[cyan downward triangles,][]{roederer2015}, NGC~6426 \citep[red upward triangles,][]{hanke2017}, and NGC~6809 \citep[purple squares,][]{Rain_2019}.}
 \label{fig:alpha_baeu}
\end{figure}

\begin{figure}
 \includegraphics[width=\linewidth]{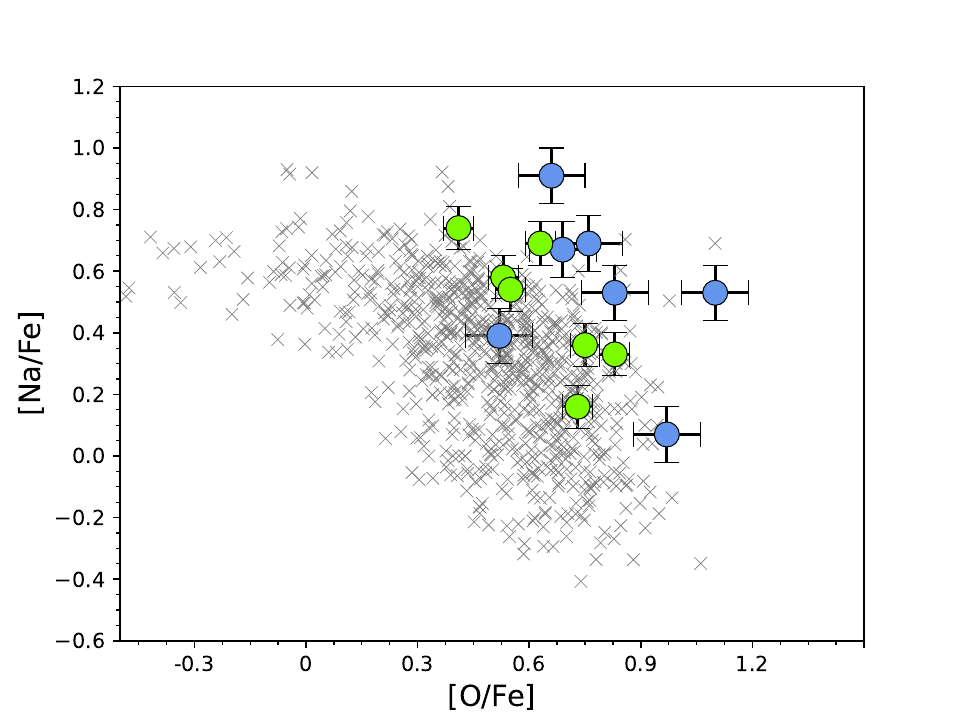}
 \caption{[Na/Fe] vs [O/Fe]. Green filled circles indicate our data. Light blue filled circles correspond to the study of \citetalias{sanroman}. Gray symbols are Galactic GC \citep{carr2009b}.}
 \label{fig:nao}
\end{figure}

The $\alpha$-elements (O, Mg, Si, Ca, and Ti) listed in Table~\ref{tab:abun} exhibit a clear overabundance relative to solar values. Since Oxygen is sensitive to the Na-O anticorrelation, its behavior is discussed in further detail in Section~\ref{nao}. As shown in Fig.~\ref{fig:alpha}, the $\alpha$-elements levels in NGC~4372 are typical for this type of system, with our measurements aligning well with those of field stars and other GCs at similar metallicities \citep[e.g. NGC~6809, NGC~4833, NGC~6426; ][respectively]{Rain_2019,roederer2015,hanke2017}.

Si has the largest observed dispersion among the $\alpha$-elements (Table~\ref{tab:errors}). Specifically, the observed dispersion ($\sigma_{obs}=0.16$) is larger than the total expected error ($\sigma_{total}=0.07$), providing evidence for an intrinsic internal scatter beyond the measurement uncertainties. This is further exemplified by one star being 0.18 dex more Si-rich than the average. This behavior is consistent with findings in other massive globular clusters where Silicon shows abundance variations potentially linked to high-temperature proton-capture processes \citep{milone17}.

Using our Mg, Si, Ca, and Ti abundances, we obtained a mean $\alpha$-element enrichment of {[$\alpha$/Fe]}= 0.33 $\pm$ 0.05. This value is in good agreement with the mean of [$\alpha$/Fe]=0.37 $\pm$ 0.07 reported by \citetalias{sanroman} (who excluded star $\#7$, identified as an outlier in their analysis). In Fig.~ \ref{fig:alphaprom}, we place our sample within the broader context of $\alpha$-element behavior across both Galactic and extragalactic objects at different metallicities. It is evident that NGC~4372 follows the same trend as other Galactic \gc of similar metallicity \citep[e.g. NGC 4833, NGC 6426, NGC 6809;][]{roederer2015,hanke2017,Rain_2019}. We have also included the four stars from \citetalias{sanroman} for which Si measurements were available. Since Ti can be synthesized through multiple nucleosynthesis channels and Mg is susceptible to depletion via proton-capture reactions, Si and Ca are often considered "pure" $\alpha$-elements. For these, we find a mean value of [(Si+Ca)/Fe]=0.38, which is slightly higher than, yet consistent with, the overall $\alpha$-average reported above.\\
These $\alpha$-elements are primarily synthesized in Type II supernovae (SNe II), which are highly efficient at producing these species \citep{sneden2004}. Consequently, the observed $\alpha$-enrichment is interpreted as a lack of Type Ia supernovae (SNe Ia) contributions to iron production within the proto-cluster cloud at these very low metallicities. To further explore potential correlations between pure SN~II enrichment and eventual time delays from neutron star mergers (assuming the latter drive the early $r$-process), we present the $[\alpha/\text{Fe}]$ vs $[\text{Ba}/\text{Eu}]$ plane in Fig.~\ref{fig:alpha_baeu}. Overall, our targets exhibit consistent $\alpha$-enhancement coupled with low, $r$-process-dominated $[\text{Ba}/\text{Eu}]$ ratios, closely tracking the distribution of Galactic halo field stars and other ancient GCs. For most $\alpha$-elements (Mg, Si, Ca, and Ti), we observe flat distributions across $[\text{Ba}/\text{Eu}]$ with no statistically significant correlations. This general absence of a systematic trend suggests that both core-collapse products and neutron-capture elements were injected rapidly and homogeneously into the proto-cluster gas during its earliest formative stages. Interestingly, however, a visual inspection of the top panel ([\text{O}/\text{Fe}] vs [\text{Ba}/\text{Eu}]) hints at a potential negative trend. A Spearman rank correlation test yields a strong negative coefficient of $\rho = -0.739$, though with marginal statistical significance ($p = 0.058$) likely owing to the limited sample size for oxygen determinations. If confirmed with a larger sample, such a tentative anticorrelation would be consistent with a time-delay scenario, wherein the onset of delayed iron production (from SNe Ia) or delayed $s$-process enrichment (from AGB stars) gradually depresses the $[\text{O}/\text{Fe}]$ ratio relative to the early $r$-process-dominated material.

According to the study of \citet{horta2020}, in-situ low-metallicity clusters display a [Si/Fe] plateau with a mean value of $\sim 0.25$, while the accreted subgroups (those belonging to the Gaia Enceladus, Helmi streams and Sequoia, for example) have [Si/Fe]$\sim 0.17$ on average in the same metallicity range. Our Si measurements in NGC~4372 are in agreement with those obtained by \citet{horta2020}, though our values are slightly higher, and support the scenario that NGC 4372 is an in-situ cluster.

\subsection{Na-O anticorrelation}\label{nao}
Variations in light element abundances (e.g., C, N, O, Na, Mg, and Al) are a ubiquitous feature among \gc and constitute some of the strongest evidence for the existence of MP \citep[see, e.g.,][and references therein]{2009a,rev2012,bastianlardo18}. In this regard, NGC~4372 is no exception. 

Regarding Na and O abundance, an analysis of star-to-star uncertainties (Table~\ref{tab:errors}) reveals that the observed spread in these elements significantly exceeds the expected analytical errors. Specifically, for Na, we find an observed dispersion of $\sigma_{obs, Na}=0.2$ dex, which is substantially higher than our mean total uncertainty of $\sigma_{tot, Na}=0.07$ dex. Similarly, for O, the observed spread of $\sigma_{obs, O}=0.14$ dex far surpasses the total error budget $\sigma_{tot, O}=0.04$ dex. This discrepancy ($\sigma_{obs} \gg \sigma_{tot}$) confirms the presence of an intrinsic chemical spread, reinforcing the existence of multiple stellar populations within NGC 4372.
As shown in Fig.~\ref{fig:nao}, the sample exhibits the well-established Na-O anticorrelation typical of nearly all Galactic \gc studied to date \citep{2009a}. While this feature was previously reported by \citetalias{sanroman}, our Na-O anticorrelation is narrower and quantitatively confirmed through a Spearman rank correlation analysis, yielding $\rho=-0.786$ with a significance of $P=0.036$. Furthermore, an Orthogonal Distance Regression (ODR) fit, which accounts for uncertainties in both axes, results in a slope of $-0.69\pm0.20$. These statistical constraints provide a more robust definition of the chemical evolution patterns in NGC 4372 compared to previous qualitative assessments. For comparison purposes, we also include Galactic \gc from \citet{2009a,Rain_2019,roederer2015}, as well as halo field stars in Fig.~\ref{fig:nao}. It is evident that our sample follows the general trend observed in other Galactic \gc. 

The Na-O anticorrelation is considered the primary evidence for the presence of MPs \citep{carretta2010}, identifying stars that belong to two (or more) stellar generations \citep{2009a, rev2012}: a FG characterized by O-rich and Na-poor stars, and a second generation (SG) consisting of stars depleted in O and enhanced in Na. As shown in Fig.~\ref{fig:nao} we identified three Na-poor/O-rich stars (those with [Na/Fe]<0.4) belonging to the primordial (FG) component, while the remaining four stars belong to the second generation. Therefore, we confirm that NGC~4372 hosts at least two distinct stellar populations.
This anticorrelation reflects material processed through proton-capture nucleosynthesis via the CNO and Ne-Na cycles, which deplete oxygen (and C) while enhancing sodium (at the expense of Ne). 

\subsection{Mg-Al anticorrelation and other light elements}\label{mgal}

In addition to the CNO and Ne-Na cycles, the Mg-Al cycle also participates in the chain of proton-capture reactions in H-burning at high temperature, above 70 MK \citep{prantzsos06}, forming Al at the expense of Mg. 

According to \citet{carr2009b}, the Mg-Al anticorrelation is predominantly found in massive and/or metal-poor \gc. In an analysis of ten Northern Hemisphere \gc using APOGEE DR10 spectra, \citet{meszaros2015} found that the Mg-Al anticorrelation became more evident as the sample size per cluster increased, although it remained absent in some cases. This is in full agreement with the broader census by \citet{pancino2017}  who confirmed that this chemical signature is not a universal feature of all \gc. 

\citetalias{sanroman} reported no significant spread in Mg abundances, primarily due to the large uncertainties in their measurements. In this work, we provide reliable Al abundances for five stars in our sample. For the remaining two stars, both of which are Na-poor, only upper limits for Al could be determined. We find that neither Mg nor Al exhibits a significant intrinsic dispersion, although Al shows a slightly larger observed spread than Mg ($\sigma_{obs}=0.14$ and $0.09$ dex, respectively). Given their chemical enrichment, it is likely that all stars with reliable Al measurements in our sample belong to the second generation (SG) of the cluster. 

\begin{figure}
 \includegraphics[width=\columnwidth]{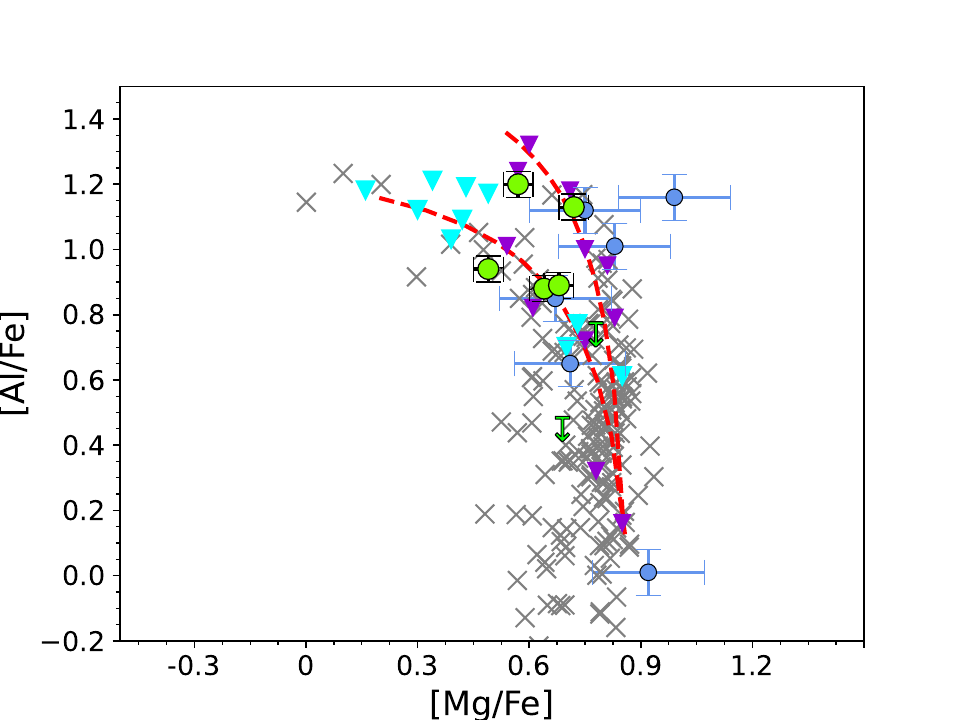}
 \caption{[Al/Fe] vs [Mg/Fe]. Green filled circles indicate our measurements, while green arrows are upper limits of Al abundances for those two stars. Light blue filled circles are data from \citetalias{sanroman}, gray symbols are different \gc \citep{carr2009b}. We also include abundances of metal-poor \gc: NGC~4833 \citep[cyan triangles,][]{roederer2015}, NGC~6809 \citep[violet triangles,][]{Rain_2019}. Red dashed lines are dilution models according to \citet{carr2009b}.}
 \label{fig:mgal}
\end{figure}

\citet{Rain_2019} identified a bimodal Mg-Al anticorrelation in NGC~6809 characterized by two distinct dilution model trajectories (following \citet{2009a}), which suggest the contribution of different polluters to the cluster's chemical enrichment. Since NGC~4273 shares similar mass and metallicity, and given that we employed the same methodology and linelist as \cite{Rain_2019}, a direct comparison is highly informative. As shown in Fig.~\ref{fig:mgal}, our sample appears to follow a similar chemical trend. Specifically, the five stars with reliable Al measurements align closely with one of the two enriched sequences found in NGC~6809, supporting the presence of a double Mg-Al anticorrelation also in NGC~4372. This distribution suggests that NGC~4372 may host a complex multiple population structure consistent with a double dilution model, although a larger sample would be required to definitively confirm this bimodal behavior.

The presence of a dual anticorrelation pattern (Na-O and Mg-Al) is not universal among globular clusters, as the Mg-Al cycle requires significantly higher temperatures (>70 MK) than the NeNa chain \citep{rev2012}. The detection of such patterns in NGC~4372, similar to those found in massive clusters like NGC~2808 \citep{carr2009b} and NGC~6752 \citep{yong03}, provides critical constraints on the nature of the polluters. The bimodal behavior suggested by our data and the comparison with \cite{Rain_2019} could imply a complex enrichment history, potentially involving different classes of polluters. This double signature reinforces the idea that NGC~4372 underwent extreme nucleosynthesis conditions during its early formation stages, placing it among the group of clusters with the most extensive chemical complexity.

\begin{figure}
 \includegraphics[width=\columnwidth]{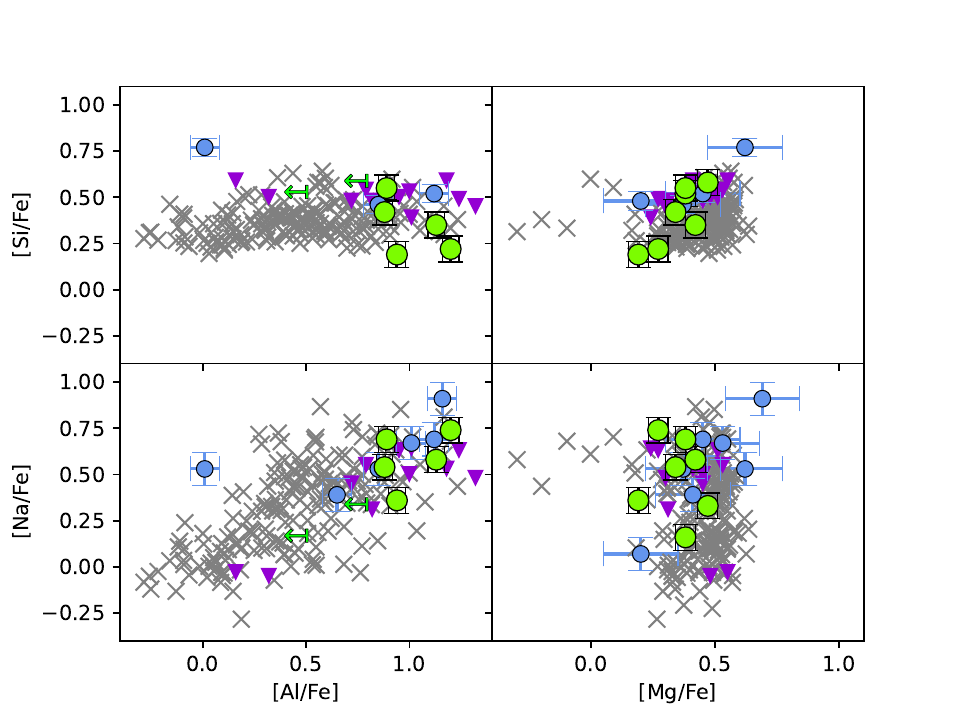}
 \caption{Top:[Si/Fe] vs [Al/Fe] and [Mg/Fe]. Bottom: [Na/Fe] vs [Al/Fe] and [Mg/Fe]. Green circles (arrows) are our data, light blue circles are from \citetalias{sanroman} while gray cross symbols are GCs from \citet{carr2009b}. Violet triangles are from \citet{Rain_2019}.}
 \label{fig:naal}
\end{figure}

Additionally, NGC~4372 shows a clear Na-Al correlation (bottom left panel of Fig.~\ref{fig:naal}).
Si is the third element involved in the Mg-Al cycle and, in our sample, exhibits a slightly larger dispersion than the other two elements ($\sigma_{obs}=0.16$). 
Specifically, we find a clear positive correlation between Si and Mg (top right panel of Fig.~\ref{fig:naal}), while a hint of anticorrelation is observed between Si and Al (top left panel). This behavior is consistent with that reported for NGC 6809 by \cite{Rain_2019}. Notably, our results differ from the findings of \citet{carr2009b}, who observed that very Al-rich stars often show an enhancement in Si. Our results suggest that the nucleosynthesis conditions in NGC~4372 did actually reach the extreme temperatures required to significantly activate the Si leakage from the Mg-Al chain.

According to theoretical models, a positive Na-Al correlation is expected when both the Ne-Na and Mg-Al chains are active. This trend is indeed observed in our sample, as shown in the bottom-left panel of Fig.~\ref{fig:naal}, where Al and Na abundances appear correlated. In contrast, no clear relationship is found between Na and Mg (bottom-right panel of Fig.~\ref{fig:naal}), a result that is consistent with the findings of \citet{Rain_2019}.

\subsection{Heavy elements}

\begin{figure}
 \includegraphics[width=\columnwidth]{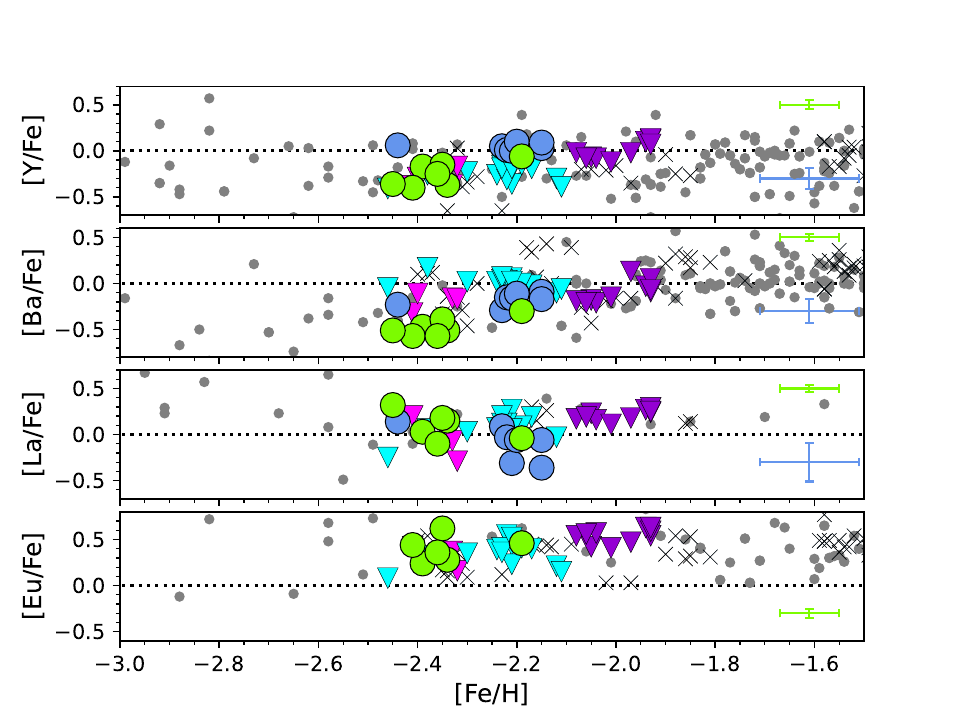}
 \caption{The neutron-capture elements [Y/Fe], [Ba/Fe], [La/Fe] and [Eu/Fe] as a function of [Fe/H]. Green filled circles indicate our targets, light blue filled circles are data from \citetalias{sanroman}, gray symbols are different \gc. We also include abundances of metal-poor GCs NGC 4833 \citep[cyan triangles,][]{roederer2015}, NGC 6426 \citep[magenta triangles,][]{hanke2017}, NGC 6809 \citep[violet triangles,][]{Rain_2019}.}
 \label{fig:heavy}
\end{figure}

\begin{figure}
 \includegraphics[width=\columnwidth]{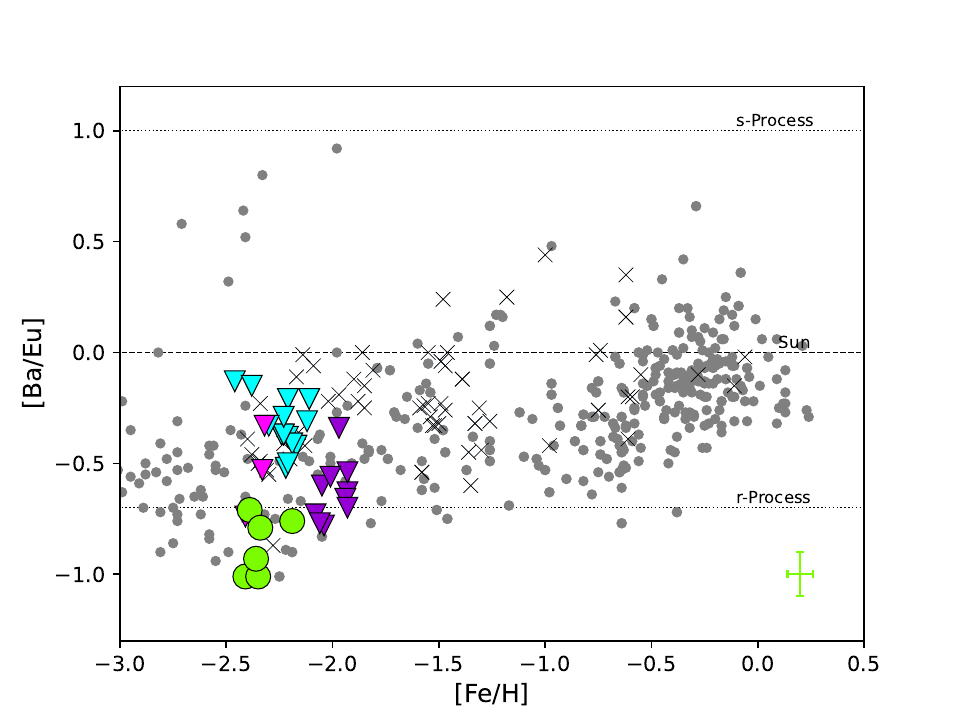}
 \caption{[Ba/Eu] vs [Fe/H]. Green filled circles indicate our targets; black symbols are different \gc. We also include abundances of metal-poor GCs NGC 4833 \citep[cyan triangles,][]{roederer2015}, NGC 6426 \citep[magenta triangles,][]{hanke2017}, NGC 6809 \citep[violet triangles,][]{Rain_2019}. Gray points are halo and disk field stars \citep{full2000,venn2004,ishi}.}
 \label{fig:baeu}
\end{figure}

Elements heavier than the iron peak (Z > 30) are primarily synthesized via neutron-capture (n-capture) processes, as proton-capture reactions become inefficient due to the increasing Coulomb barrier. These processes are traditionally classified into two distinct channels based on the relative timescale of neutron addition versus $\beta$-decay: the slow $s$-process and the rapid $r$-process, where neutron capture occurs on timescales much shorter than $\beta$-decay lifetimes. In this work, we present abundance measurements for six n-capture elements: Y, Ba, La, Ce, Nd, Eu. While $s$-process nucleosynthesis is predominantly associated with the thermally-pulsing AGB phase of low-to-intermediate mass stars \citep{karakas2014, gallino98}, the astrophysical site of the $r$-process remains a subject of active debate. Traditionally linked to core-collapse supernovae SNe II, although recent evidence identifies neutron star mergers as a major production site, contributing significantly to the abundance of Eu \citep{abbot2017}.

The abundance ratios of the n-capture elements analyzed in this work are presented in Fig.~\ref{fig:heavy}, alongside data for field and globular cluster stars for context. For comparison, we overplot the results from \citetalias{sanroman}, who reported a small apparent dispersion in Y, La, and Ba abundances, although an intrinsic spread could not be confirmed due to observational uncertainties. Notably, our [Y/Fe] values are subsolar, showing a systematic underabundance that aligns with results for other metal-poor GCs such as NGC~4833 and NGC~6426, but contrast with the nearly solar ratios found in NGC~6809.
Furthermore, our [Ba/Fe] ratios are more depleted than those reported by \citetalias{sanroman} and other metal-poor \gc \citep[e.g., NGC 6426, NGC 4833, NGC 6809;][]{roederer2015, hanke2017, Rain_2019}, yet they remain consistent with the overall trend observed in field stars. In contrast, La is slightly overabundant, while Eu is significantly supersolar with a mean value of <[Eu/Fe]>=0.40$\pm 0.6$. This strong $r$-process enrichment is in excellent agreement with the abundance patterns typically observed in metal-poor GCs (see discussion \cite{rev2004}). 

In addition, we measured abundance ratios for two other n-capture elements: Neodymium (Nd) and Cerium (Ce). We find a mean [Nd/Fe] value of 0.03 dex and [Ce/Fe] value of $-$0.17 dex, with observed dispersions of $\sigma_{\rm obs}=0.15$ and $\sigma_{\rm obs}=0.13$, respectively. These mean values are consistent with other GCs of similar metallicities \citep[e.g.][]{hanke2017,Rain_2019}. While the dispersion in Ce is comparable to our measurement uncertainties ($\sigma_{tot}=0.13$ dex), the spread in Nd is significantly larger than its associated error ($\sigma_{tot}$ = 0.05 dex vs $\sigma_{obs}=0.15$ dex), suggesting a potential intrinsic spread. However, this Nd variation shows no significant correlation with the light element Na. This decoupling indicates that the enrichment of the n-capture elements in NGC~4372 was independent of the nucleosynthetic process responsible for the MPs phenomenon. Instead, it likely reflects primordial inhomogeneities in the proto-cluster gas, where the $r-$ or $s-$ process material was not fully homogenized before the onset of star formation.

In Fig.~\ref{fig:baeu}, we show the [Ba/Eu] ratio as a function of metallicity. This ratio is a powerful diagnostic tool, as it quantifies the relative contributions of the $s$- and $r$-process to the chemical evolution of the cluster. Notably, this work provides the first measurements of Eu for NGC~4372, filling a gap in the previous chemical characterization of this system. In the case of NGC 4372, we found a mean [Ba/Eu] value of $-0.87$, indicating that the nucleosynthesis of n-capture elements was dominated exclusively by the $r$-process. The determination of Eu is crucial since previous studies identified an enrichment in heavy elements; the lack of an $r$-process tracer made it difficult to disentangle the enrichment timescales. Our results confirm that the cluster formed from gas primarily enriched by core-collapse supernovae (which also explains the $\alpha$-element enrichment) before low-mass AGB stars could contribute significantly to the chemical inventory through the $s$-process.

\section{Conclusions}\label{conclusions}
In this work, we performed a detailed chemical abundance analysis of seven RGB stars in the metal-poor GC NGC~4372. Using high-resolution optical spectra from FLAMES/UVES, we derived the abundances of 17 chemical elements. Specifically, four elements were measured using the EW method, while the remaining 13 were analyzed via spectral synthesis. Furthermore, we conducted a rigorous error analysis that enabled a direct comparison with the study of \citetalias{sanroman}, despite our samples being entirely independent. Our analysis significantly extends the chemical characterization of the cluster by providing the first measurements of Zn, Ce, Eu, and Nd, elements that were not included in the study by \citetalias{sanroman}. The addition of these species, particularly the $r$-process tracer Eu and the lanthanides Ce and Nd, allows for a more comprehensive assessment of the neutron-capture nucleosynthesis history in this metal-poor system. While one star in our sample is 0.16 dex more metal-rich than the average, it exhibits no other chemical or physical peculiarities. Our main findings are as follows:

\begin{itemize}
\item NGC 4372 exhibits a mean metallicity of [Fe/H]=$-2.36 \pm 0.03$ with an observed dispersion of $\sigma_{\rm obs}=0.03$ dex, in excellent agreement with previous studies. Furthermore, we ruled out the presence of an intrinsic metallicity spread within the cluster.
\item The iron-peak elements generally show good agreement with the trends observed in other Galactic GCs and halo field stars, with no detectable abundance spread. 
\item The $\alpha$-elements show an enhancement of [$\alpha$/Fe]=0.33$\pm$ 0.05 dex, in good agreement with the results of \citetalias{sanroman}. This enrichment is typical for Galactic GCs at this metallicity.
\item Our [Si/Fe] measurements are consistent with the mean value proposed by \citet{horta2020}, suggesting that NGC 4372 is an in-situ GC. This reinforces the classification of this cluster as a native member of the Galactic Halo or Thick Disc.
\item We detect a potential intrinsic abundance spread in Silicon. The observed dispersion in [Si/Fe] ($\sigma_{obs}=0.16$) is significantly larger than the total expected uncertainties ($\sigma_{tot}=$0.07), suggesting that NGC 4372 might not be chemically homogeneous regarding $\alpha-$elements. This aligns with other massive and chemically complex Galactic GCs (eg., NGC 6752, NGC 2808), where Si variations are often linked to extreme proton-capture processing. Specifically, this implies that the Mg-Al chain 'leaked' into Si production via ${}^{27}\text{Al}(p, \gamma){}^{28}\text{Si}$, a process that requires temperatures exceeding $\sim$ 64 MK. Consequently, our results suggest that NGC 4372 hosted a highly energetic progenitor population (likely massive AGB stars or FRMS) capable of reaching the thermal regimes necessary to bypass standard Mg-Al cycle limits, reinforcing its classification as a chemically complex cluster.
\item We confirm that NGC~4372 hosts multiple populations (MPs), as evidenced by a well-defined Na-O anticorrelation (see Fig.~\ref{fig:nao}). Our analysis provides a clearer detection of this chemical signature compared to the results of \citetalias{sanroman}, reinforcing the classification of this cluster as a complex stellar system with at least two distinct generations.
\item Our identification of a double Mg-Al anticorrelation in NGC 4372 represents a novel finding that suggests a complex enrichment history. While this bimodal signature appears quite clearly in our current analysis, a larger sample size is essential to definitively confirm its presence. Confirming this feature is of significant importance, as it would provide critical evidence for discrete pollution episodes and offer new constraints on the formation scenarios of metal-poor GCs.
\item The [Ba/Eu] ratios in NGC 4372 consistently fall below [Ba/Eu] < 0, revealing that the heavy-element content is dominated by the $r$-process. This indicates that the progenitor gas was enriched primarily by high-energy events, such as magnetorotational supernovae or neutron star mergers, before the onset of significant $s$-process contributions from low-mass AGB stars. The dominance of the $r$-process at this metallicity underscores the rapid formation scale of NGC 4372, suggesting that the cluster’s initial chemical footprint was locked in by massive, short-lived progenitors in a poorly mixed proto-Galactic environment.
\end{itemize}

\section*{Data availability}
The data underlying this article are already public and available in the ESO archive page \url{http://archive.eso.org/eso/eso_archive_main.html} under the program ID 088.B-0492(A). \\

\begin{acknowledgements}
     C.B. acknowledges financial support from Fondecyt regular grant 1251882. S.V. gratefully acknowledges the support provided by Fondecyt reg. 1220264, the ANID BASAL project FB210003.
 C.M. thanks the support provided by  ANID-GEMINI  Postdoctorado No.32230017.
\end{acknowledgements}

\bibliographystyle{aa} 
\bibliography{biblio.bib}

\end{document}